\documentclass[twocolumn]{aastex63}
\usepackage{graphicx}

\submitjournal{AJ}
\shorttitle{A New Proper Motion Determination of Leo~I}
\shortauthors{Casetti-Dinescu et al.}

\begin{document}


\title{A New Absolute Proper Motion Determination of Leo~I Using HST/WFPC2 images and Gaia 
EDR3}

\correspondingauthor{Dana I. Casetti-Dinescu}
\email{casettid1@southernct.edu,dana.casetti@gmail.com}

\author[0000-0001-9737-4954]{Dana I. Casetti-Dinescu}
\affiliation{Department of Physics, Southern Connecticut State University, 501 Crescent Street, 
New Haven, CT 06515}
\affiliation{Astronomical Institute of the Romanian Academy, Cutitul de Argint 5, Sector 4, 
Bucharest, Romania}
\author{Caitlin K. Hansen}
\affiliation{Department of Physics, Southern Connecticut State University, 501 Crescent Street, 
New Haven, CT 06515}
\author{Terrence M. Girard}
\affiliation{Department of Physics, Southern Connecticut State University, 501 Crescent Street, 
New Haven, CT 06515}
\affiliation{Department of Astronomy, Yale University, Steinbach Hall, P.O. Box 208101, New 
Haven, CT 06520-8101}

\author{Vera Kozhurina-Platais}
\affiliation{Space Telescope Science Institute, Baltimore, MD 21218}

\author{Imants Platais}
\affiliation{Department of Physics and Astronomy, The Johns Hopkins University, Baltimore, MD 
21218}


\author[0000-0003-2159-1463]{Elliott P. Horch}
\affiliation{Department of Physics, Southern Connecticut State University, 501 Crescent Street, 
New Haven, CT 06515}

\begin{abstract}
We measure the absolute proper motion of Leo~I using a WFPC2/HST data set that
spans up to 10 years, to date the longest time baseline utilized for this satellite.
The measurement relies on $\sim 2300$ Leo~I stars
located near the center of light of the galaxy;
the correction to absolute proper motion is based on 174
{\it Gaia} EDR3 stars and 10 galaxies.
Having generated highly-precise, relative proper motions for all {\it Gaia} EDR3
stars in our WFPC2 field of study, our correction to the absolute EDR3 system
does {\bf not} rely on these {\it Gaia} stars being Leo~I members. 
This new determination also benefits from a
recently improved astrometric calibration of WFPC2.
The resulting proper-motion value,
$(\mu_{\alpha}, \mu_{\delta}) = (-0.007\pm0.035,-0.119\pm0.026)$ mas yr$^{-1}$
is in agreement with recent, large-area, {\it Gaia} EDR3-based determinations.
We discuss all the recent measurements of Leo~I's proper motion and adopt
a combined, multi-study average of
$(\mu_{\alpha}^{3meas}, \mu_{\delta}^{3meas}) = (-0.036\pm0.016,-0.130\pm0.010)$ mas yr$^{-1}$.
This value of absolute proper motion for Leo~I indicates its
orbital pole is well aligned with that of the Vast Polar Structure,
defined by the majority of the brightest dwarf-spheroidal satellites of the
Milky Way.
\end{abstract}


\section{Introduction} \label{sec:intro}
Highly precise and accurate proper-motion measures of distant Milky Way satellites remain 
difficult to come by in spite of the
spectacular progress made by ESA's {\it Gaia} mission. The distances to these satellites
places their stars at the faint end of the {\it Gaia} measurements where the proper-motion
uncertainties are largest, making these systems challenging. 
Nevertheless, there are important science drivers for improving the state of proper-motion 
measurements for such systems as recently pointed out
by \citet{pawlowski2020milky}. One aspect highlighted by this study is the polar alignment of 
the most massive Milky Way (MW) satellites: the more accurate the proper motions, the more 
apparent this alignment is. The co-orbiting of the most massive MW satellites is an 
observational aspect that challenges state-of-the-art cosmological simulations 
\citep{pawlowski2020milky}.

Although \citet{pawlowski2020milky} include {\it Gaia} DR2 proper motions \citep{gaia2018gaia} 
in their work, two distant satellites stand out as having large uncertainties in the position of 
their orbit poles: Leo~I and Leo~II (see their Figure 1). 
Notably, Leo~I's pole is outside the 
area where other satellites' poles cluster. Clearly, better measurements are needed for these 
two satellites.

In this study we will focus on Leo~I. 
Our new proper-motion determination makes use of archival HST WFPC2 
exposures spanning up to 10 years and {\it Gaia} EDR3 \citep{gaiacollaboration2020gaia} to 
obtain the correction to absolute proper motion. A new distortion calibration of WFPC2 
\citep{casettiwfpc2} is utilized. The WFPC2 exposures cover a small field of view very near the 
center of light of Leo~I.
This ensures that we only sample Leo~I stars near the center of mass of the system. One 
advantage of this is to alleviate any offsets due to possible proper-motion gradients within
Leo~I's internal kinematics over the field of study. 
Proper-motion studies of such satellites that rely directly on {\it Gaia} measurements
of satellite stars are susceptible to such effects
due to the trade off between the shallowness of {\it Gaia} and the areal coverage needed to 
obtain a reasonable number of member stars.

Radial velocity studies \citep{Koch2007, Sohn2007, Mateo2008} do not find a significant spatial 
gradient in the line-of-sight velocity field for Leo~I. 
Of course, this does not preclude the existence
of a gradient in the tangential velocity field.
Given the large areas over which
candidate members are selected in {\it Gaia}-based proper-motion studies of
dwarf galaxies, \citep[$\sim r_{tidal}$ or more, see e.g.,][]{Martinez-Garcia2021}, we can estimate
a rough upper limit as to the expected proper-motion gradient due to a plausible velocity gradient.
To do so, we assume one of the largest gradients found in the Hercules system \citep{Aden2009}, 
specifically 16 km~s$^{-1}$kpc$^{-1}$.
At the distance of Leo~I, this corresponds to a delta of 0.012 mas yr$^{-1}$ 
for a spatial offset of $r_{tidal}$ in the field of Leo~I.
This value approaches some of the proper-motion
errors quoted for Leo~I in the recent literature.

\section{Literature Review of Leo~I Proper-Motion Determinations} \label{sec:litrev}
\begin{figure*} 
\includegraphics[scale=0.64,angle=-90]{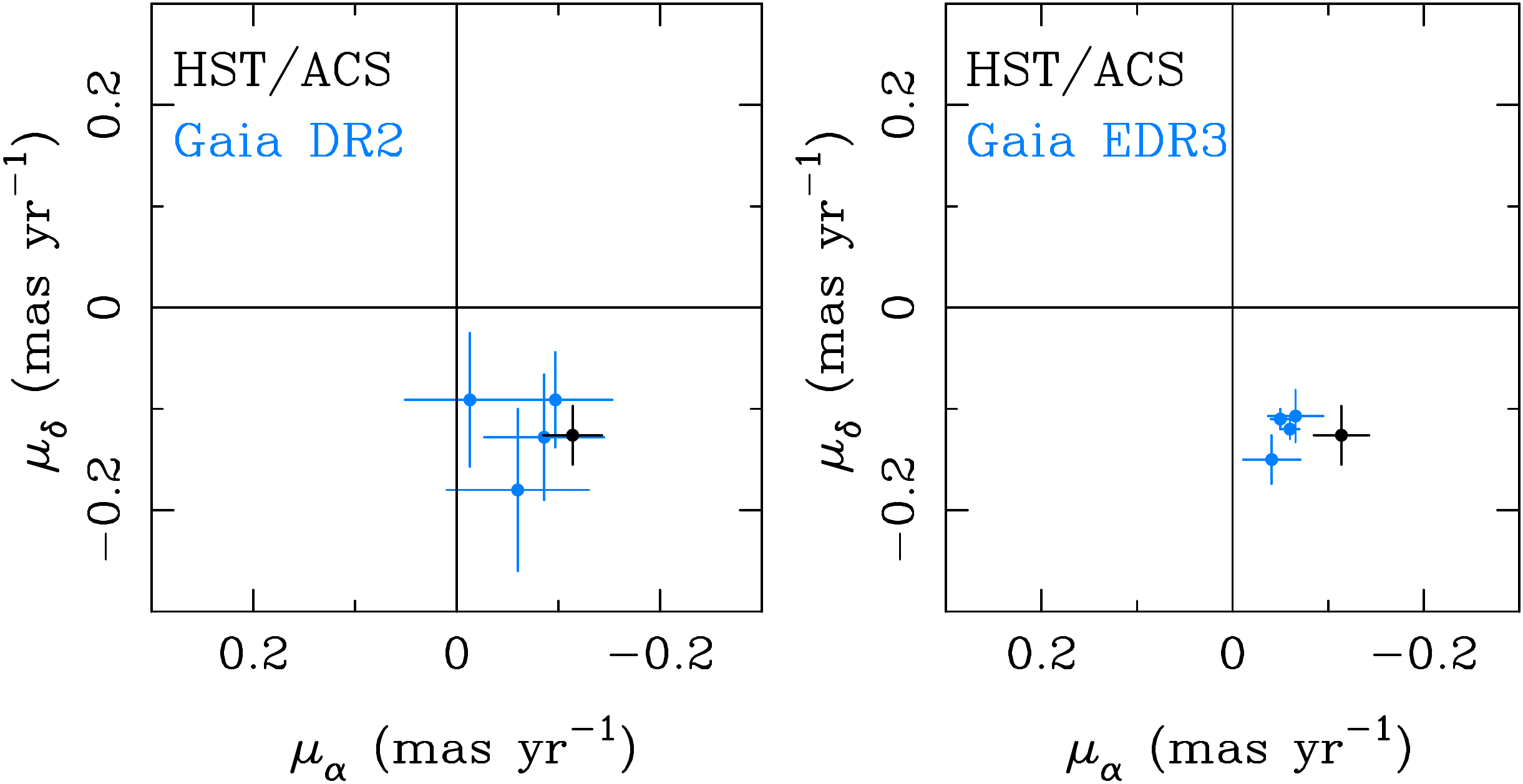}
\caption{Absolute proper-motion determinations for Leo~I. Besides the 2013
  HST ACS/WFC determination, the left panel shows {\it Gaia} DR2-based measurements and 
the right panel shows those based on {\it Gaia} EDR3.
Differences within the {\it Gaia} DR2 and EDR3 groups are primarily the
manner in which Leo~I members were selected.
See also Table \ref{tab:leo1studies}.}
\label{fig:pm_status}
\end{figure*}

Previous measurements for Leo~I are summarized in Table \ref{tab:leo1studies} and Figure 
\ref{fig:pm_status}.
In Tab. \ref{tab:leo1studies} we present the reference, the data source, the proper motion 
values with formal uncertainties, the time baseline, the approximate number of Leo~I stars and 
the approximate number of absolute proper-motion calibrators used in the determination. 
Calibrators can be {\it Gaia} stars or extragalactic sources. For completeness, we also include 
our study's determination. 
Among the differences between the various {\it Gaia} DR2 or EDR3 studies are the ways in which 
Leo~I members were selected: either spectroscopically, as for most studies, or via a 
combination of properties yielding a membership probability 
\citep{mcconnachie2020revised,McConnachie2020,Battaglia2021}.

In Fig. \ref{fig:pm_status} we present the 2013 HST determination by \citet{sohn2013space} 
compared to the {\it Gaia} DR2 (left panel) and {\it Gaia} EDR3 (right panel) determinations. 
The average value of the {\it Gaia} DR2 measurements is
$(\mu_{\alpha}, \mu_{\delta}) = (-0.064\pm0.019, -0.122\pm0.021)$ mas yr$^{-1}$, where the 
uncertainties are from the standard deviation of the four 
measurements\footnote{Throughout the paper 
$\mu_{\alpha}$ actually represents $\mu_{\alpha}$ cos $ \delta$, and units are mas~yr$^{-1}$.}. 
Specifically, the standard deviation is $(\sigma_{\mu_{\alpha}}, \sigma_{\mu_{\delta}}) = 
(0.037, 0.042)$ mas yr$^{-1}$, and can be interpreted as being due to the different ways in which 
Leo~I members were selected.
Otherwise, results should have been identical since they all use {\it Gaia} DR2 stars. 
Formal uncertainties for each {\it Gaia} DR2 determination are actually larger than the scatter 
indicates as they should also include any systematic errors in {\it Gaia} DR2.
Since the average {\it Gaia} DR2 value is not obtained from independent measurements, it is 
best to assume its uncertainty is of the order of the standard deviation, or $\sim 0.04$ mas 
yr$^{-1}$. In this case,
the average {\it Gaia} DR2-based value is consistent with the 2013 HST measurement.

The {\it Gaia} EDR3-based determinations have smaller formal errors than the DR2 ones, 
with an average value
$(\mu_{\alpha}, \mu_{\delta}) = (-0.054\pm0.006, -0.122\pm0.010)$ mas yr$^{-1}$, and a standard 
deviation
$(\sigma_{\mu_{\alpha}}, \sigma_{\mu_{\delta}}) = (0.011, 0.020)$ mas yr$^{-1}$. The average 
value agrees very well with the
{\it Gaia} DR2-based average, while the scatter is reduced by a factor of between two and four. 

\begin{deluxetable*}{lccccrr} 
\tablecaption{Leo~I Proper-Motion Determinations \label{tab:leo1studies}}
\tablewidth{0pt}
\tablehead{
    \colhead{Reference} &
    \colhead{Data Sources} &
    \colhead{$\mu_{\alpha}\cos(\delta)$} &
    \colhead{$\mu_{\delta}$} &
    \colhead{Time Baseline} &
    \colhead{$N_{Leo~I}$} &
    \colhead{$N_{ZP}$}
}
\startdata
\cite{sohn2013space} & ACS/WFC  & -0.114 $\pm$ 0.029 & -0.126 $\pm$ 0.029 & 5 years& $\sim$  
36000 & $\sim$ 100 \\ \\
\cite{gaia2018kinematics} & Gaia DR2   & -0.097 $\pm$ 0.056 & -0.091 $\pm$ 0.047 &  22 months& 
174 & 174\\
\cite{fritz2018pm} & Gaia DR2  & -0.086 $\pm$ 0.059 & -0.128 $\pm$ 0.062 & 22 months & 241 & 
241 \\
\cite{simon2018gaia} & Gaia DR2  & -0.013 $\pm$ 0.064 & -0.091 $\pm$ 0.066 & 22 months & 187 & 
187 \\
\cite{mcconnachie2020revised} & Gaia DR2  & -0.060 $\pm$ 0.070 & -0.180 $\pm$ 0.080 & 22 months 
& 15 & 15 \\ \\
\cite{McConnachie2020} & Gaia EDR3 & -0.050 $\pm$  0.010 & -0.110 $\pm$  0.010  & 34 months  & 
15$^a$ & 15$^a$ \\
\cite{Li2021} & Gaia EDR3 & -0.066 $\pm$  0.029 & -0.107 $\pm$  0.026  & 34 months  & 368 & 368 
\\
\cite{Martinez-Garcia2021} & Gaia EDR3 & -0.041 $\pm$  0.030 & -0.150 $\pm$  0.024  & 34 months 
& 294 & 594$^b$ \\
\cite{Battaglia2021} & Gaia EDR3 & -0.060 $\pm$  0.010 & -0.120 $\pm$  0.010  & 34 months & 
1342 & 1342 \\ \\
this study           &  WFPC2 + EDR3 & -0.007 $\pm$  0.035 & -0.119 $\pm$  0.026  & 10 years & 
$\sim$ 2300 &  174+10$^c$ \\ \\
\enddata
\tablecomments{All proper-motion determinations given in mas yr\textsuperscript{-1}. Column 1 
lists the reference, Column 2 the source of the data, Column 3 and 4 give the proper motion in 
each coordinate with corresponding uncertainty estimates. In all but one of the Gaia 
determinations these estimates do not 
include potential systematic errors. Column 5 gives the time baseline, Column 6 gives the 
number of stars in Leo~I considered members in the respective study (or the number of stars 
with membership probability $> 50\%$ for the \cite{mcconnachie2020revised} and 
\cite{Battaglia2021}) studies, 
Column 7 gives the number of objects (galaxies/QSOs/{\it Gaia} stars) 
used to derive the correction to absolute proper motion, i.e., zero point. \\
  $a$ - \cite{McConnachie2020} do not specify the number of Leo~I members; we assumed the same 
number as in their previous 2020 study. \\
  $b$ - This study performs a local correction to absolute proper motion
  using QSOs within 3 degrees of Leo~I's center, in
  an attempt to minimize potential systematic errors. Formal uncertainty estimates 
also include the effects of 
systematic errors. \\
  $c$ - Our study uses 174 {\it Gaia} stars and 10 galaxies.}
\end{deluxetable*}

The one
result somewhat discrepant from the other three is that of \citet{Martinez-Garcia2021} who 
readjusts the absolute proper motion by using local (within a radius of 3 degrees) QSOs, 
rather than relying on the EDR3 direct measurements of Leo~I stars. This is done in an attempt 
to eliminate any local systematic errors.
Assuming once again an uncertainty in the
{\it Gaia} EDR3 average value based on the standard deviation, the EDR3 value is found
inconsistent with the
2013 HST measurement in $\mu_{\alpha}$ at a $2\sigma$ level. 
A discrepancy of this size
can alter the location of the orbital pole in a substantial way (see Section \ref{sec:pole}).
\vspace{5mm}

\section{HST WFPC2 Data Set} \label{sec:wfpc2data}

The WFPC2 served as the principal imaging instrument on the HST from 1993 to 2009,
accumulating more than 135,000 exposures (see WFPC2 Data Handbook v10.0, MAST\footnote{Barbara Mikluski
Archive for Space Telescopes, https://archive.stsci.edu.}).
The WFPC2 consists of four different detectors, three of which are nearly identical. The three 
Wide Field Cameras at f/12.9 provide an ``L" shaped field of view at $2.5\arcmin \times 
2.5\arcmin$ with each 15 $\mu$m detector pixel subtending $0.10\arcsec$ on the sky. The 
Planetary Camera at f/28.3 provides a field of view of $35\arcsec \times 35\arcsec$ with each 
pixel subtending $0.046\arcsec$
\citep[see WPFC2 instrument handbook v10.0,][]{astrometry3}.

The camera was deemed challenging for astrometry due to its undersampled CCDs \citep{and2000} 
as well as due to large optical distortion \citep{anderson2003} and charge transfer 
efficiency (CTE) effects \citep{dol2009}. Nevertheless, the WFPC2 archive offers a long time 
baseline for proper-motion studies, although this potential has yet to be fully exploited. 
Our team started a project to astrometrically recalibrate this instrument utilizing all 
suitable images in filters F555W, F606W, and F814W in combination with {\it Gaia} EDR3 
\citep{casettiwfpc2}, thus enabling
the possibility of new proper-motion studies using the WFPC2 archive (see below in 
\ref{sec:processing}).
To this end, we searched the archive for a relevant science target represented by repeated 
exposures over a long time baseline, exposures that would include a reasonable number of 
{\it Gaia} stars. We decided to investigate the Leo~I 
set of exposures, the properties of which are listed in
Table \ref{tab:dataset_prop}. There are 12 F555W exposures separated by $\sim 5$ years and 16 
F814W exposures separated by $\sim 10$ years.

\begin{deluxetable}{lccc} 
\tablecaption{WFPC2 data set properties \label{tab:dataset_prop}}

\tablewidth{0pt}
\tablehead{
    \colhead{PID} &
    \colhead{Filter} &
    \colhead{$N_{exp} \times T_{exp}$(sec)} &
    \colhead{Epoch}
}
\startdata
 5350 & F555W & $3 \times 1900$ & 1994.17 \\
 5350 & `` & $1 \times 350$ & `` \\
 8095 & F555W & $1 \times 500$ & 1999.46 \\
 8095 & `` & $7 \times 400$ & `` \\
\hline
 5350 & F814W & $3 \times 1600$ & 1994.17 \\
 5350 & `` & $1 \times 300$ & `` \\
 9817 & F814W & $4 \times 600$ & 2004.44 \\
 9817 & `` & $4 \times 500$ & `` \\
 9817 & `` & $4 \times 300$ & `` \\
\enddata
\end{deluxetable}

\begin{figure*}
    \centering
    \includegraphics[scale=0.50, trim=0in 2.4in 0in 0 in]{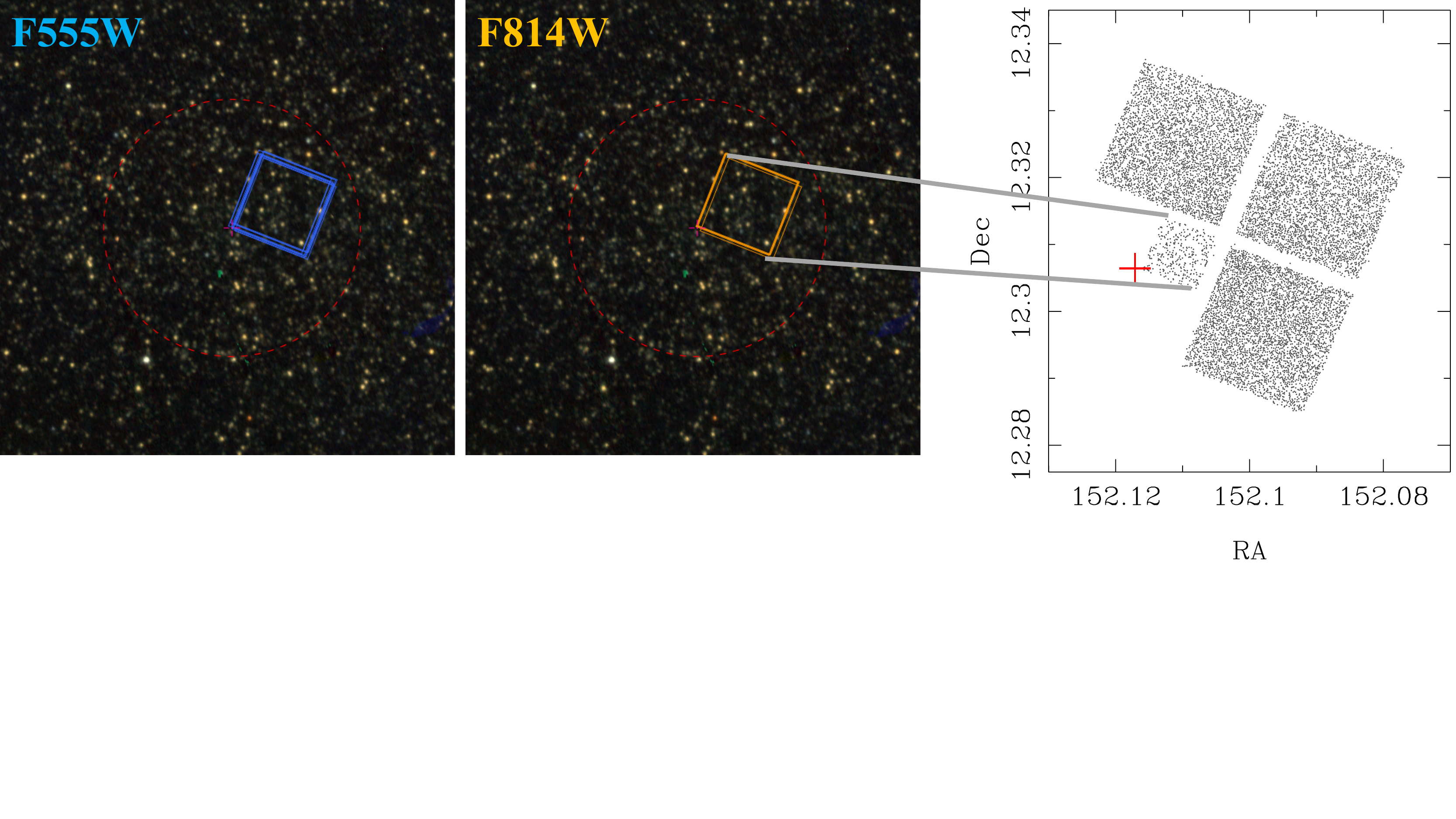}
    \caption{The field of view of our data set. The PC footprint of all F555W (left) and F814W 
(middle) exposures is highlighted with the PANSTARRS survey in the background. The right 
panel displays the entire field of view as represented by our resulting proper-motion catalog. 
The nominal center 
of Leo~I is marked with a red cross in all panels. A circle of radius $60\arcsec$ is 
indicated in the left and middle panels for scale.
      For comparison, the half-light radius of Leo~I is $198\arcsec$. All exposures have the 
same orientation, and offsets are of the order of a few arcseconds.}
    \label{fig:footprint}
\end{figure*}

In Figure \ref{fig:footprint} we show the field of view of our data set. Left and middle panels 
show the PC footprint of all F555W and F814W exposures respectively, with the PANSTARRS 
survey as background. The right panel shows the entire field of view as given by objects in 
our F555W proper-motion catalog. Leo~I's center, adopted from \citet{mcconnachie2020revised}, is 
marked with a red cross in all panels. It is seen that all exposures have practically the 
same orientation and with very small offsets, of the order of a few arcseconds.

\section{Data Processing} \label{sec:processing}
\subsection{Detection and Centering of Stars in WFPC2 Exposures} \label{subsec:centering}

The WFPC2 calibrated $_{-}$c0m.fits images are downloaded from The Mikulski Archive for Space 
Telescopes (MAST)
and split into separate chip files. Each WFPC2 chip is treated separately as an individual unit 
throughout our
procedures. As stated in the WFPC2 instrument handbook, the pixels of the PC undersample the
point spread function (PSF) by a factor of about two at visual wavelengths and the pixels of
the WF chips undersample the PSF by a factor of about four at visual wavelengths.
Therefore, we choose to employ the effective Point Spread Function (ePSF) code described
by \cite{and2000} to determine a position and flux for each object in each
exposure.
This super-resolution code was developed specifically for undersampled images and has been
shown to outperform
other centering algorithms \citet{casettiwfpc2}. The method uses an empirically-determined PSF
that is fit to target images. For our purposes, the PSF used in the fitting process is based on 
a unique set of dithered HST/WFPC2 images combined into an ePSF library.  Clearly,
the ePSF library is best-suited for observations near the epoch of the data set used
to construct it; this was early in the lifetime of WFPC2 on the HST. 
We note that this particular centering algorithm
provides very high precision centers for fainter stars, as seen empirically in
the average residuals generated by transformations into standard catalogs \citep{casettiwfpc2}.

\subsection{Astrometric corrections} \label{subsec:pre}
There are a number of sources of systematic errors in the WFPC2 astrometry that are 
well-documented and modeled.
The initial two adjustments to the WFPC2 positions are the 34th-row correction and the 
correction for nominal
cubic distortion \citep{anderson1999wfpc2, anderson2003}. 
The 34th-row correction accounts for a
small manufacturing defect on each chip that made every 34th-row of pixels about 3\% narrower.
The nominal cubic distortion accounts for optically-induced geometric distortion of each of the 
WFPC2 chips.  It is expressed as a filter-independent set of 
cubic coefficients, one set for each chip, that are listed in the WFPC2 
instrument handbook.
To further refine the WFPC2 positions, we then apply the recently determined geometric 
higher-order distortion-correction maps for each filter. 
These correction maps are based on {\it Gaia} EDR3 and all suitable
WFPC2 exposures in the appropriate filter \citep{casettiwfpc2}.

Finally, we explore CTE effects in the WFPC2 positions. To do so, we adopt a first-epoch,
long exposure as a reference exposure and transform all other exposures into it.
The transformation is a classic polynomial one using as reference stars all well-measured
stars on the CCD chip \citep[see e.g.,][and references therein]{casetti_sextans}.
The vast majority of the stars in our
exposures are Leo~I members. The intrinsic proper-motion dispersion of Leo~I stars is much
lower than that given by
position measurement errors. Therefore measurement error will dominate the scatter of the 
residuals
of the transformation, even when epochs of the exposures differ. Trends of these residuals
with magnitude should reveal systematics of the CTE type. 
When transforming an exposure near that of the reference exposure, the residuals are found
to be flat in both chip coordinates; however, when comparing different epochs, slopes in
residuals with magnitude are apparent in both $x$ and $y$ coordinates.
We monitor these slopes for each chip and exposure and find that the slopes do vary from chip
to chip, but do not change significantly from one late-epoch exposure to another.
Thus we stack residuals per chip and filter, determining mean slopes and applying them as
corrections to the positions of the late-epoch exposures. In Table \ref{tab:slopes}
we present the values of these slopes. Note that the magnitudes are uncalibrated, instrumental 
magnitudes.

The physical interpretation of these values is not straightforward: $y$-slope values are not 
consistently larger than $x$-slope values as one would expect from CTE effects and the $y$ 
direction being the readout direction (Recall
that the orientation of all exposures is similar; see Sect.~\ref{sec:wfpc2data}). 
Furthermore, the 
length of the exposure time did not make a difference in the values of the slopes. It appears 
that the epoch difference is the only factor
that is relevant.  The slopes are not very different between filters, although keep in mind 
the F814W set has
twice as long a time baseline as the F555W set. Slopes also differ between individual WF chips. 
Possibly, the ePSF
changed substantially between early-90s and late-90s observations, and this is what the
slopes reflect;
although we cannot pinpoint what caused the ePSF change, be it CTE and/or some other effect.
The slope corrections are empirically derived, from internal comparisons of the WFPC2 data.

Once positions are corrected for these magnitude-dependent linear trends, they all are on 
the system of the early-epoch exposures.

\begin{deluxetable}{lrr}
\tablecaption{Slopes of late-epoch residuals as a function of magnitude. \label{tab:slopes}}
\tablewidth{0pt}
\tablehead{
    \colhead{Chip} &
    \colhead{$x$-slope} &
    \colhead{$y$-slope} \\
    \colhead{} &
    \colhead{(mpix/mag)} &
    \colhead{(mpix/mag)}
}
\startdata
& F555W & \\
\hline
PC & $-10.8 \pm 1.7$       &  $ -9.5  \pm 1.4 $   \\
WF2 & $2.9  \pm 0.6$       &   $ 3.3  \pm 0.6 $   \\
WF3 & $8.6  \pm 0.8$       &   $-4.8  \pm 0.8 $   \\
WF4 & $-2.9  \pm 0.7$      &   $-7.1  \pm 0.7 $   \\
\hline
& F814W & \\
\hline
PC & $-7.5 \pm 1.1$         & $-9.8  \pm 1.1$   \\
WF2 & $ 0.0  \pm 0.5$       &    $3.6  \pm 0.6$  \\
WF3 & $7.0  \pm 0.7$        &   $-2.2  \pm 0.8$   \\
WF4 & $-2.0  \pm 0.6$       &   $-9.6  \pm 0.7$
\enddata
\end{deluxetable}

\begin{figure*}
  \epsscale{1.15}
\plottwo{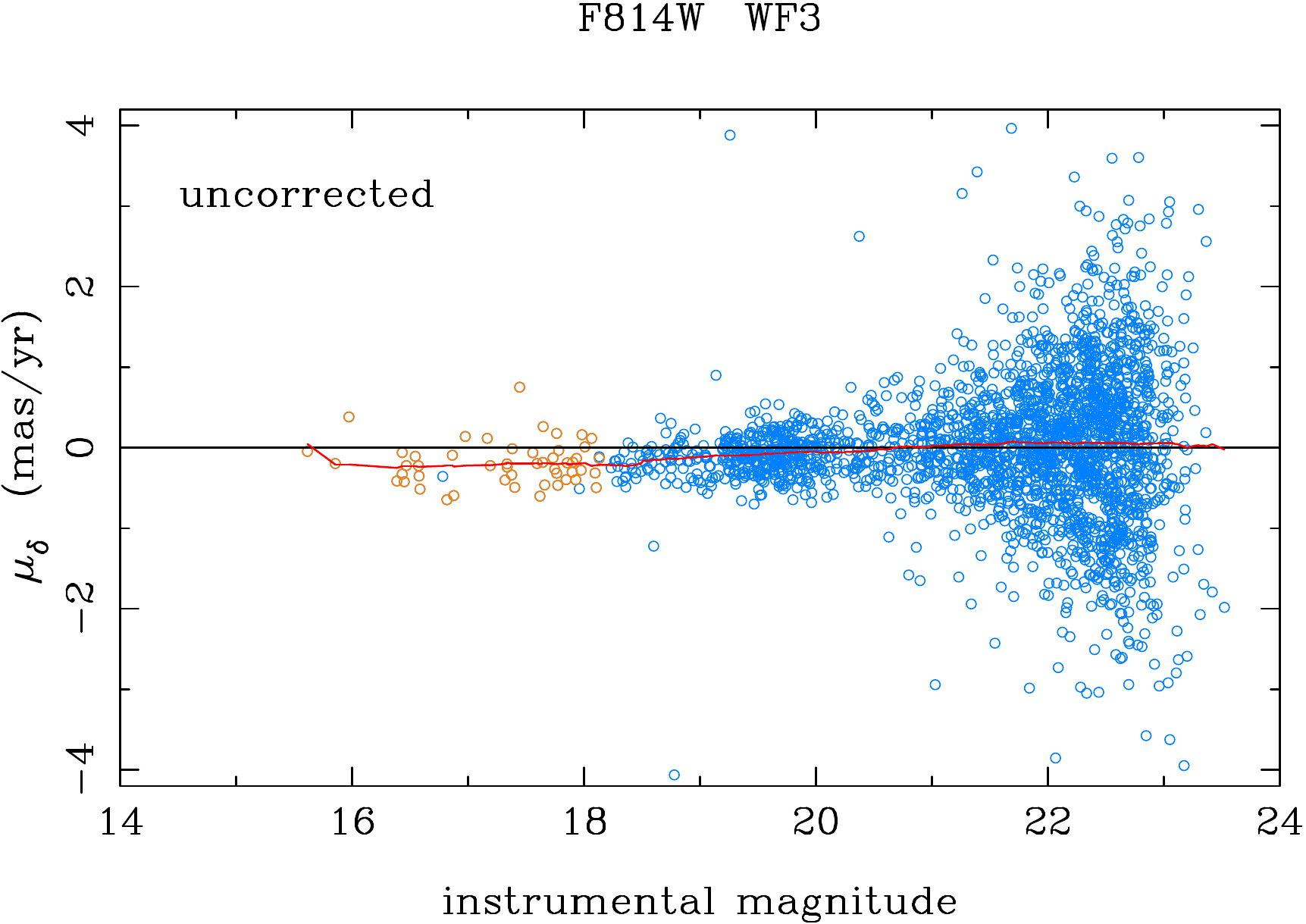}{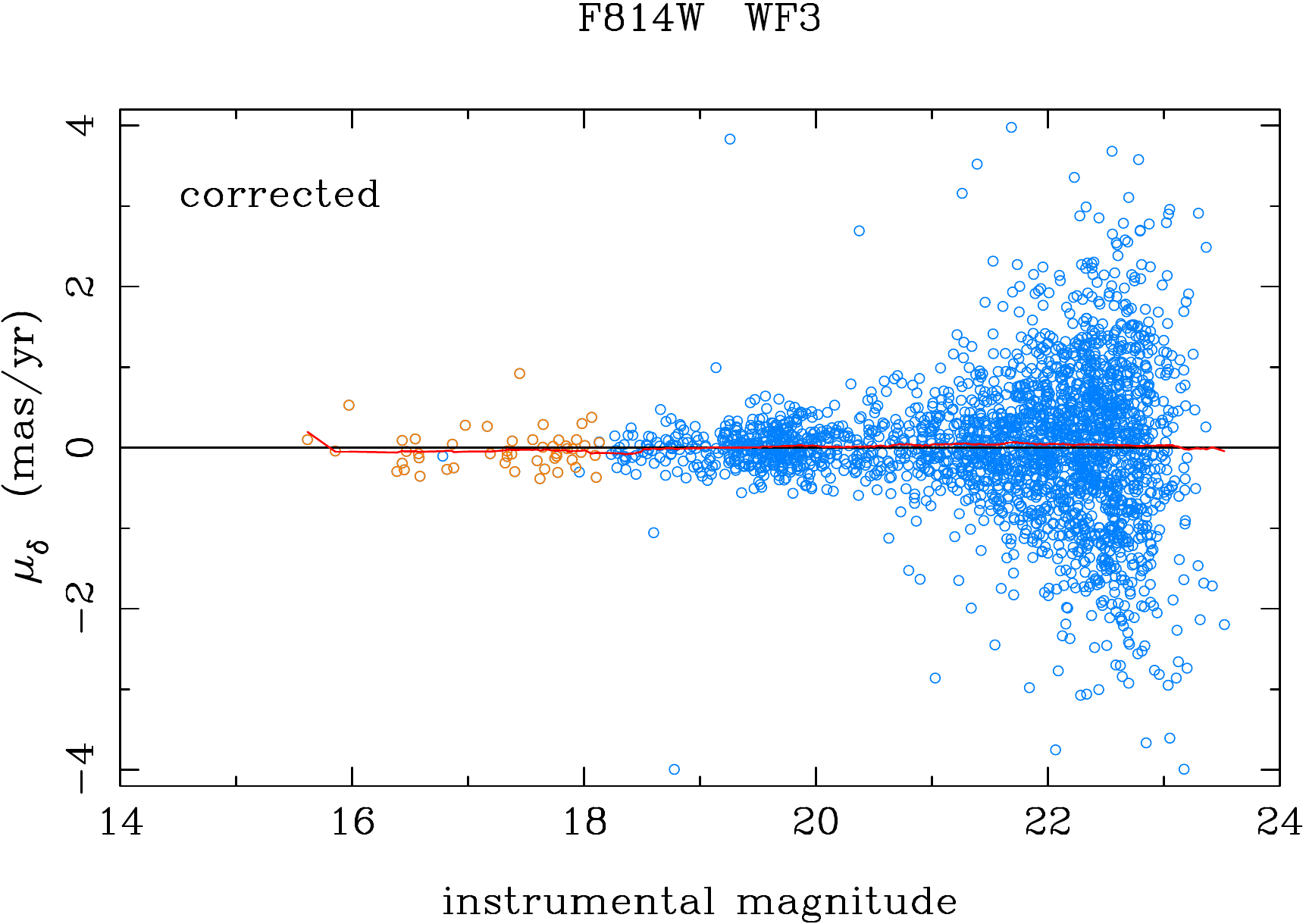}
\caption{Relative proper motions, $\mu_{\delta}$, as a function of magnitude for filter F814W, 
chip WF3. {\bf Left} panel shows the {\bf uncorrected} values, {\bf right} panel, those 
derived after {\bf correcting} the late-epoch positions for the linear magnitude dependence. 
{\it Gaia} stars are highlighted with orange symbols.
The red curve shows the magnitude trend via a moving mean.
Similar plots were constructed and inspected 
for all chips, in component $\mu_{\alpha}$ as well, and in filter F555W.
Each exhibited roughly similar behavior.
The $G$-magnitude range of the {\it Gaia} stars shown here is 
between 18.4 and 21.}
\label{fig:mag_cor}
\end{figure*}

To check the impact of our magnitude-dependent correction we calculate relative proper motions
(see also Sect.~\ref{sec:pmdet}), rotate them into the celestial coordinate
(RA,Dec) system using {\it Gaia} stars, and plot these as a function of magnitude. 
If the corrections 
are appropriate, no trend with magnitude should be visible in the proper motions. In Figure 
\ref{fig:mag_cor} we show one such example, where
the uncorrected proper motions are shown in the left panel, while the corrected ones are in 
the right panel.
{\it Gaia} stars at the bright end are highlighted. After inspecting many such plots we 
conclude that the relative proper motions are largely free of any systematic trend. We note 
that the magnitude-slope corrections were derived (and applied)
using residuals expressed in chip coordinates
while the proper motions are in a rotated (RA,Dec) system.
The degrees to which these proper-motion plots are flat with respect to magnitude is an
indication that the correction was done adequately.

\section{Proper-Motion Determination} \label{sec:pmdet}

\subsection{Relative Proper Motions} \label{subsec:rel}

The WFPC2 frames must be transformed to a common system in order to measure their proper motions 
differentially.
We start by aligning the WFPC2 exposures in a detector-based coordinate system using a 
classical
plate-overlap solution \citep[e.g.,][]{casetti_sextans,casettiwfpc2}. We utilize two 
early-epoch exposures,
a 1900-sec exposure in F555W and a 1600-sec exposure in F814W,
as our reference exposures. The assumption is that these exposures are least
affected by CTE \citep[e.g.,][]{dol2009}.

The transformation consists of up to 3rd order polynomials.
We estimate the plate constants\footnote{These parameters are labeled as ``plate'' parameters
  to align with the historical use of the terminology when astrometry measurements were made 
with photographic plates.}
by selecting reference stars on the detector field that are well-centered and
with multiple measurements in our set of exposures. Stars with instrumental
magnitudes ranging from 10 to 21 best fit these criteria.
We use an iterative least squares procedure to refine the plate constants and the 
relative proper motions;
in the initial iteration we assume zero proper motions for the reference stars.

Besides the WFPC2 exposures, we
include in the procedure the {\it Gaia} EDR3 {\bf positions}. 
The original {\it Gaia} EDR3 celestial coordinates (RA, Dec)
are roughly transformed into
detector coordinates \citep{casettiwfpc2}, and then treated as another
``exposure'' to the set of exposures for a given filter.
In effect, the faint {\it Gaia} stars (G = 18.4 - 21) will benefit from having new
relative proper motions derived from an increased baseline of
$\sim 20$ years time difference, yielding very high precision proper motions,
of the order of 0.1 mas yr$^{-1}$ for the F814W set.
This is an advantage that can be exploited in other, future relative proper-motion studies.

We determine proper motions for all objects with at least four position measurements
separated by a 3 year minimum.
We perform six iterations of the least-squares procedure with outlier culling at 2.5$\sigma$
for the reference stars.
Formal proper-motion uncertainties are calculated from the scatter
about the best fit line in the position versus time plots.
This process produces {\bf relative} proper motion measurements in detector coordinates 
with
units of millipixels per year.
The output positions and proper motions are then converted to the celestial coordinate system 
(RA, Dec) via a 
transformation using the {\it Gaia} stars on
each chip

\begin{figure*}
\includegraphics[scale=0.68,angle=-90]{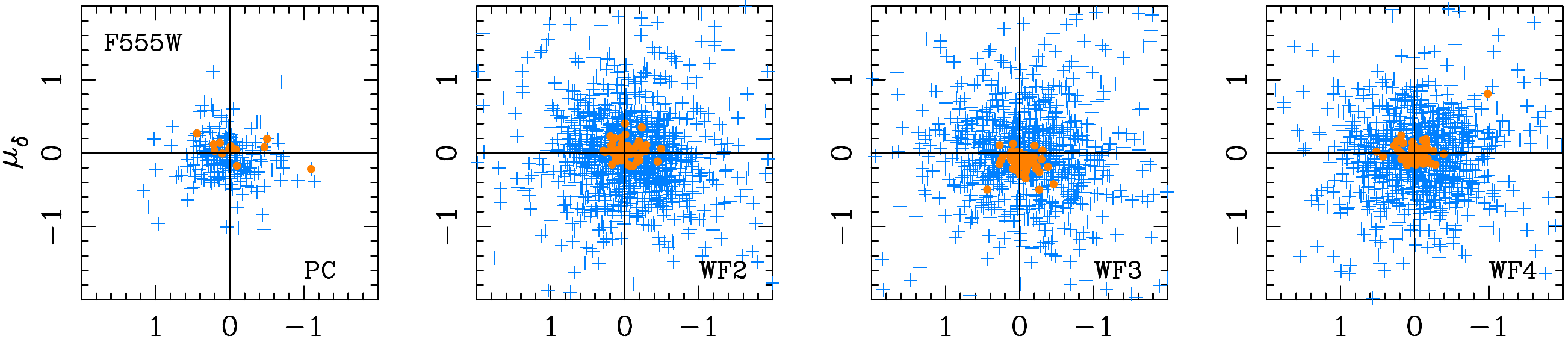}
\includegraphics[scale=0.68,angle=-90]{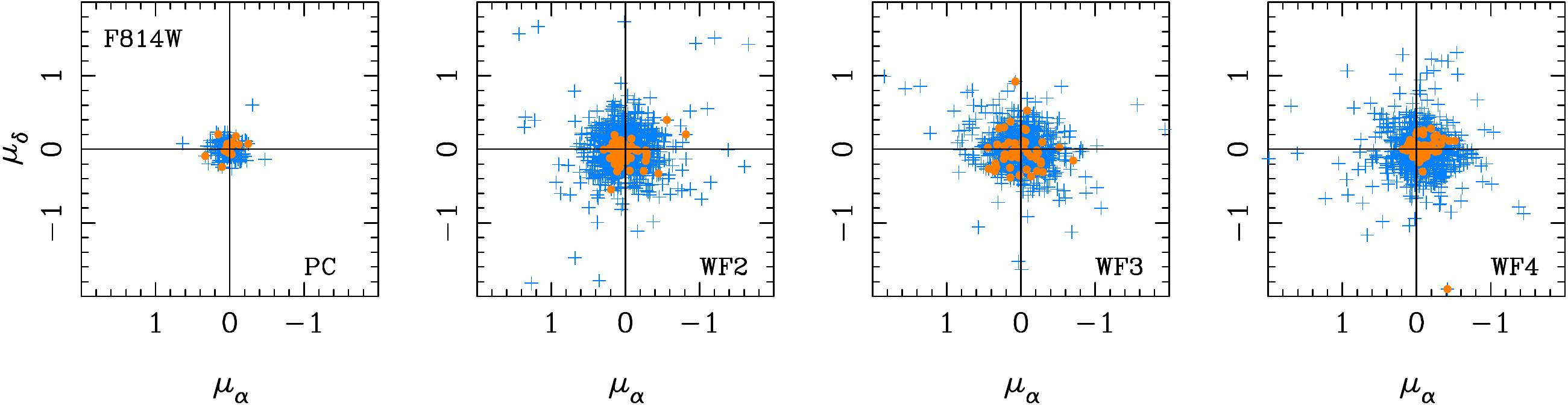}
\caption{Relative proper-motion diagrams per chip and filter for objects with instrumental 
magnitude $< 21$.
  {\it Gaia} stars are highlighted with round orange symbols. Units are mas yr$^{-1}$.
  F814W proper motions of non-{\it Gaia} stars are much tighter than those of F555W
  due to their having a 10-yr baseline compared to the 5-yr baseline for F555W.}
\label{fig:rel_pm_vpd}
\end{figure*}

The stars that participated in the plate solution are predominantly Leo~I stars for the 
following reasons:
the field is very near the center of mass of Leo~I (see Fig. \ref{fig:footprint}),
the Galactic latitude of this field is rather high at $49\arcdeg$,
the magnitude range is dominated by faint stars ($G > 18$), thus
excluding foreground Galactic stars, and finally, the proper motion
dispersion is very low (see Fig. \ref{fig:mag_cor})
indicating a kinematically cold population, unlike the foreground Galactic stars.
Therefore, our system of relative proper motions is roughly tied to the system of the 
Leo~I dwarf spheroidal galaxy itself.
In Figure \ref{fig:rel_pm_vpd} we show the relative proper motions for well-measured stars, 
highlighting the
{\it Gaia} stars.

Next, we identify the Leo~I stars that will be used in the determination of the average
relative motion of the system. The selection of Leo~I stars to include in this average is 
straightforward
given that the target field is likely dominated by Leo~I stars as we argued above.
We calculate a simple average of the relative proper motions for each chip and filter.
Outliers are removed based on the following criteria:
instrumental magnitude $>$ 21 in both filters, total proper motion error calculated as the 
square root of the squared values in each coordinate larger than 0.4 mas yr$^{-1}$ and 0.8 mas 
yr$^{-1}$ in the PC and WF respectively for filter F555W, and larger than 0.25 mas yr$^{-1}$ 
and 0.5 mas yr$^{-1}$ in the PC and WF respectively for filter F814W. After these cuts, we also 
remove stars with proper motions larger
than 3$\sigma$ from the average in an iterative fashion.
Corresponding uncertainties in
these averages are determined from the standard deviations of the measured proper motions.
The average relative proper motion as determined per chip and per filter is presented in Table 
\ref{tab:pm_relative},
where we have specified the number of stars used in each determination.
Due to the smaller field of view of the PC compared to that of the WF chips,
fewer stars are used in the PC solutions.
The values in Tab. \ref{tab:pm_relative} are not far from zero, as expected, since the relative
proper-motion solution used a 
frame of reference largely composed of Leo~I stars.
Still, the calculation of a precise, relative proper-motion value for a properly cleaned
sample of Leo~I stars is important.

\begin{deluxetable*}{c|rrc|rrc}
\tablecaption{Relative Proper-Motion Values of Leo~I per chip and filter 
\label{tab:pm_relative}}
\tablewidth{0pt}
\tablehead{
    {Chip} &
    \multicolumn{3}{c}{F555W} &
    \multicolumn{3}{c}{F814W} \\
&
    $\mu_{\alpha,relative}$ &
    $\mu_{\delta,relative}$ &
    \# of Stars &
    $\mu_{\alpha,relative}$ &
    $\mu_{\delta,relative}$&
    \# of Stars
    }
\startdata
PC1 & $0.040 \pm 0.021$ & $-0.026 \pm 0.018$ & 192  & $0.013 \pm 0.009$ & $-0.001 \pm 0.008$ & 
149  \\
WF2 & $-0.036 \pm 0.016$ & $ 0.001 \pm 0.015$ & 853 & $0.009 \pm 0.009$ & $-0.015 \pm 0.009$ & 
724 \\
WF3 & $-0.066 \pm 0.021$ & $-0.007 \pm 0.019$ & 586 & $-0.005 \pm 0.011$ & $-0.013 \pm 0.009$& 
568  \\
WF4 & $-0.058 \pm 0.018$ & $-0.035 \pm 0.016$ & 684 & $-0.065 \pm 0.010$ & $-0.006 \pm 0.009$& 
592
\enddata
\end{deluxetable*}

\subsection{{\it Gaia}-based Absolute Proper Motion} \label{subsec:Gaia-based}

To determine the proper motions with respect to an inertial reference frame, 
i.e., absolute proper motions, we make use of {\it Gaia} EDR3 stars.
We calculate differences between the {\it Gaia} EDR3 absolute proper motions and our 
relative proper motions for {\it Gaia} stars that fall on each chip, in each filter.
The weighted average of these differences is used to determine the proper-motion offset,
$\mu_{offset}$, with separate offsets derived for each chip and filter.
During the procedure, we eliminate outliers via an iterative 3$\sigma$ clipping about a simple 
average,and then perform the weighted average.
The weights are based on the square of the quadrature sum of the
{\it Gaia} EDR3 catalog proper-motion uncertainties and the formal uncertainties of our 
relative proper motions. 
The {\it Gaia} errors dominate,
with average values of the order of 0.75 mas yr$^{-1}$ in RA and 0.53 mas yr$^{-1}$ in Dec. 
Comparatively, our relative proper motion uncertainties are on average 0.1 mas yr$^{-1}$ in 
both coordinates. We have 
also inspected the Renormalised Unit Weight Error (RUWE) values of the {\it Gaia} EDR3 stars.
We remind the reader that RUWE values are expected to be around 1.0 for sources where the
single-star model provides a good fit to the astrometric observations, while values $> 1.4$
indicate problematic sources for the astrometric solution \citet{Lindegren2021}.
We find that the RUWE values of our sample peak at 1.02, 
and of the 174 stars, only four have values larger than 1.2.
Otherwise, we did not explicitly rely on the RUWE values in this process.

It is important to note that the {\it Gaia} stars participating in the determination
of these offsets {\it need not be} Leo~I stars, although probably the vast majority are, 
given the characteristics of this data set.
The derived offset values are listed in Table \ref{tab:pm_offsets}. 
The PC offset values have the largest 
uncertainties due to the
small number of {\it Gaia} stars and the fact that {\it Gaia} proper-motion uncertainties 
dominate the
error budget.

\begin{deluxetable*}{c|rrc|rrc}
\tablecaption{{\it Gaia}-based proper-motion offset values, i.e., corrections to the absolute 
system \label{tab:pm_offsets}}
\tablewidth{0pt}
\tablehead{
    {Chip} &
    \multicolumn{3}{c}{F555W} &
    \multicolumn{3}{c}{F814W} \\
&
    $\mu_{\alpha,offset}$ &
    $\mu_{\delta,offset}$ &
    \# of Stars &
    $\mu_{\alpha,offset}$ &
    $\mu_{\delta,offset}$&
    \# of Stars
    }
\startdata
PC1 & $0.257 \pm 0.122$ & $0.149 \pm 0.089$ & 15 & $0.055 \pm 0.126$ & $0.204 \pm 0.090 $ & 15  
\\
WF2 & $0.003 \pm 0.073$ & $-0.201 \pm 0.055$ & 52 & $-0.008 \pm 0.078$ & $-0.118 \pm 0.058$ & 
51 \\
WF3 & $0.043 \pm 0.069$ & $-0.025 \pm 0.049$ & 49 & $0.092 \pm 0.074$ & $-0.129 \pm 0.054$ & 47 
\\
WF4 & $-0.098 \pm 0.066$ & $-0.149 \pm 0.047$ & 57 & $0.034 \pm 0.075$ & $-0.238 \pm 0.054$ & 
54
\enddata
\end{deluxetable*}

\begin{figure}
\includegraphics[scale=0.45]{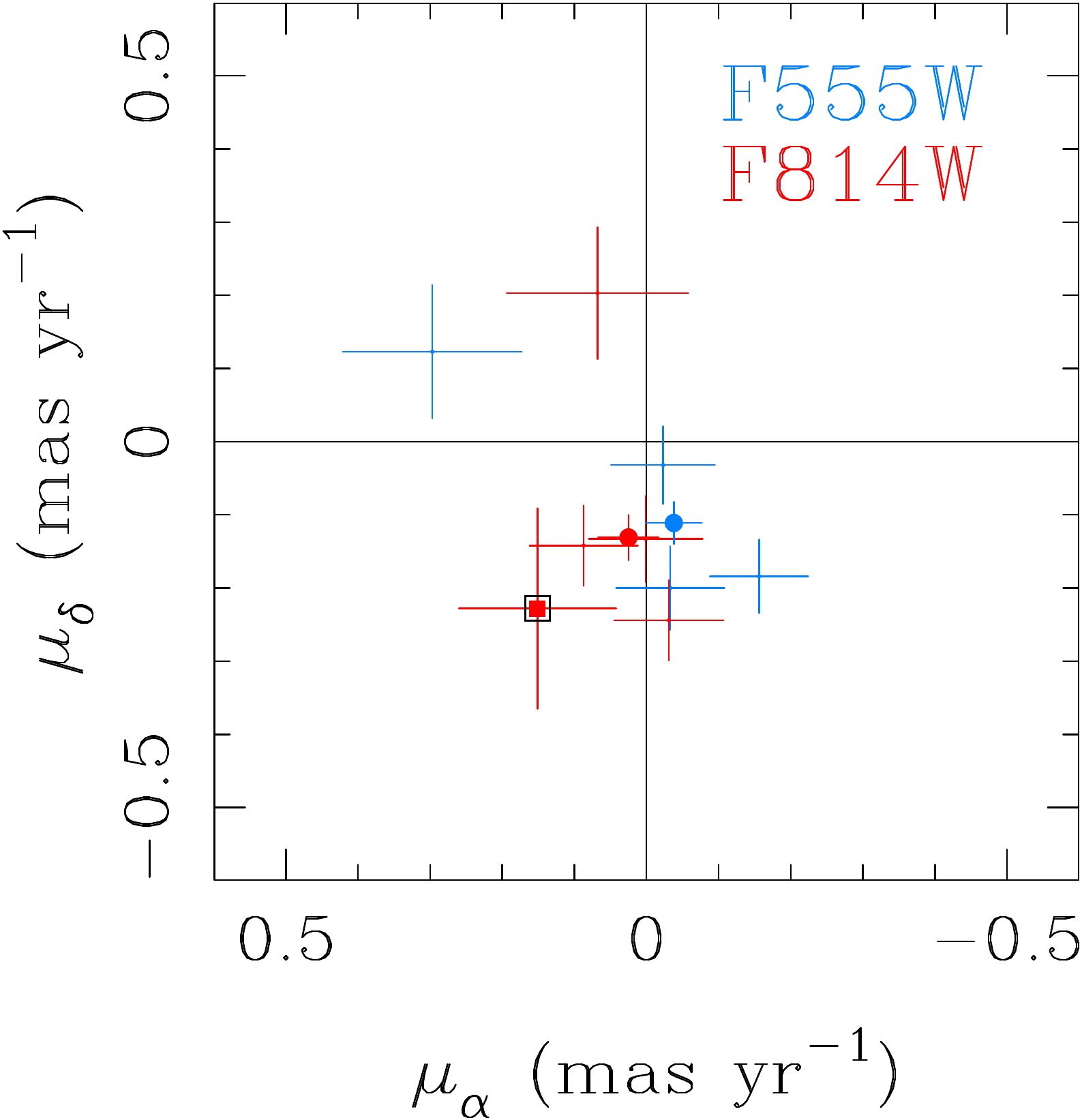}
\caption{Absolute proper-motion determinations of Leo~I. 
Our {\it Gaia} EDR3-based measurements are shown with 1$\sigma$ error bars, color coded
by filter.  Each symbol represents a WFPC2 chip, the PC values lying in the upper left
quadrant of the plot.
The error-weighted averages, per filter,
are represented by the large filled-circle symbols with smaller uncertainties. 
The square symbol shows the galaxy-based 
determination, which was
made only with F814W images.}
\label{fig:pm_abs}
\end{figure}

We apply the offsets from Tab.~\ref{tab:pm_offsets} to the average relative proper-motion 
values listed in Tab.~\ref{tab:pm_relative} to obtain
an absolute proper motion for Leo~I in each chip and filter. 
Uncertainties are given by the quadrature sum of the uncertainties in 
Tables \ref{tab:pm_relative} and \ref{tab:pm_offsets}. We list the resulting absolute 
proper-motion estimates
in Table \ref{tab:pm_abs}. For each filter, we also calculate a weighted average absolute 
proper motion across the four chips, where the weights are based on the uncertainties of each 
chip's determination; results are listed in the last line of Tab.~\ref{tab:pm_abs}.
In Figure \ref{fig:pm_abs} we show these various determinations. 
The PC estimates stand apart
from the ensemble of the WF estimates, along with their larger uncertainties, nonetheless
we choose to include them in the
per-filter weighted average, realizing they contribute only slightly 
due to the weighting scheme.
The weighted-average values for the two filters agree well, given their uncertainties.

\begin{deluxetable*}{c|rr|rr}
\tablecaption{{\it Gaia}-based absolute proper motion determinations per chip and filter 
\label{tab:pm_abs}}
\tablewidth{0pt}
\tablehead{
    {Chip} &
    \multicolumn{2}{c}{F555W} &
    \multicolumn{2}{c}{F814W} \\
&
    $\mu_{\alpha}$ &
    $\mu_{\delta}$ &
    $\mu_{\alpha}$ &
    $\mu_{\delta}$
    }
\startdata
PC1 & $0.297 \pm 0.124$ &  $0.123 \pm 0.091$    & $0.068 \pm 0.126$ &  $0.203 \pm 0.090$  \\
WF2 & $-0.033 \pm 0.075$ & $ -0.200 \pm 0.057$  & $0.001 \pm 0.079$ &  $-0.133 \pm 0.059$ \\
WF3 &  $-0.023 \pm 0.072$ &  $-0.032 \pm 0.053$  & $0.087 \pm 0.075$ &  $-0.142 \pm 0.055$  \\
WF4 & $-0.156 \pm 0.068$ &  $-0.184 \pm 0.050$  & $-0.031 \pm 0.076$ &  $-0.244 \pm 0.055$  \\ 
\hline
w.a. & $-0.038 \pm 0.039$ & $-0.111 \pm 0.029$  & $0.025 \pm 0.042$ & $-0.131 \pm 0.031$
\enddata
\end{deluxetable*}

\subsection{Galaxy-based Absolute Proper Motion} \label{subsec:galaxy-based}

Background galaxies can also be used to determine the
correction to absolute proper motion.
Our WFPC2 data set is not well-suited for identifying compact galaxy images due to the 
severe undersampling.
Nevertheless --- using only the PC images --- we median average the three deepest exposures 
from the first
epoch in each filter and inspect them visually. We find four galaxies of which only one made it 
into our relative proper-motion catalogs.
Alternatively, we make use of the galaxy classification from \citet{sohn2013space} who 
have kindly provided the necessary information for cross-matching. 
The \citet{sohn2013space} study is much deeper than ours, and also their ACS/WFC images are 
better sampled.
We match our catalogs with that of \citet{sohn2013space} and, from 
their $\sim$ 100 galaxies,
we find 10 galaxies in common
with our F555W list and 11 galaxies with our F814W list. The small number of matches is due to 
the shallowness of our data set
and the incomplete ($\sim 70\%$) areal overlap between the two studies.
Specifically, our study has a Vegamag F814W magnitude range of 19.8 to 25.6,
using the magnitude scale from
\citet{sohn2013space}. The single galaxy found
on the PC in common with the \citet{sohn2013space} study is also the one we identified by
visual inspection.

Since there are so few galaxies overall, we do not attempt per-chip estimates as in Sec. 
\ref{subsec:Gaia-based}.
Instead, we subtract from the entire set of relative proper motions the Leo~I averages from 
Tab. \ref{tab:pm_relative}, per chip, placing them all on a common system. 
Afterward, the ensemble of galaxies' proper motions will, in effect, indicate the reflex 
proper motion of the Leo~I system. 
Unfortunately, the galaxies' F555W-based relative proper-motion errors are too large
to allow a reliable zero-point determination; recall the shorter time baseline in this filter.
For the F814W data, however, we are able to determine a useful, error-weighted average 
proper motion for the galaxies, after eliminating one obvious outlier. 
The resulting value is 
$(\mu_{\alpha}, \mu_{\delta}) = (0.151\pm0.109,-0.228\pm0.137)$ mas yr$^{-1}$,
and represents
Leo~I's absolute proper motion with respect to ten galaxies measured with the F814W set. 
Although the uncertainties are large,
this measurement is consistent with the other per-chip and per-filter determinations and 
serves as a useful check on the
{\it Gaia}-based measurements.
Our galaxy-based determination is also included in Fig. \ref{fig:pm_abs}.

\section{Final Results and Comparison with Other Studies}  \label{sec:results}

To produce a final estimate of Leo~I's absolute proper motion we
must appropriately combine the {\it Gaia}-based results 
from Section \ref{subsec:Gaia-based}, one in each filter, and the single 
galaxy-based estimate of Section 
\ref{subsec:galaxy-based}.

Note that the
{\it Gaia}-based estimates per filter are not independent since their uncertainties 
are dominated by the {\it Gaia} EDR3 catalog uncertainties, while the vast majority of 
{\it Gaia} reference stars are the same in the reductions of the two filter data sets.
Specifically, there were seven {\it Gaia} stars that were in the F555W estimate and {\bf not} 
in the F814W estimate,
and there was one {\it Gaia} star that was in the F814W estimate and {\bf not} in the F555W 
estimate. 
Thus, one cannot simply do an error-weighted average of the two filter's values and 
naively propagate the errors to determine the error of the average.

To circumvent this difficulty we proceed as follows. We place all relative proper motions 
of the {\it Gaia} stars
on a common proper-motion system, that of Leo~I, by subtracting the values
in Tab. \ref{tab:pm_relative} from their
initial relative proper-motion values, in the same manner as in
Sec. \ref{subsec:galaxy-based}. 
At this point, all the stars' proper motions
will be on the same system, regardless of chip or filter. 
For each {\it Gaia} EDR3 star, we then do a weighted average of these ``Leo~I-system''
proper motions for filter F555W and F814W, and propagate through the uncertainties. 
The weights are based on the relative proper-motion uncertainties in each filter. If the star
only appears in one filter's set, we simply retain that value. 
Next, for each {\it Gaia} star, we take the difference between its ``Leo~I-system'' proper
motion and its EDR3 catalog proper motion: this will be an estimate of the absolute proper 
motion of Leo~I based on this single {\it Gaia} star.  Its uncertainty is the quadrature
sum of the relative (``Leo~I-system'') and absolute (EDR3) uncertainties.
There are a total of 174 such {\it Gaia} EDR3 stars for which proper-motion differences
can be determined.
We calculate an error-weighted average of these to produce
a single, combined-filter, {\it Gaia}-based absolute proper motion of Leo~I:
$(\mu_{\alpha}^{EDR3}, \mu_{\delta}^{EDR3}) = (-0.025\pm0.037,-0.115\pm0.027)$
mas yr$^{-1}$. 

As one would expect, this data point agrees well with the per-filter
estimates (see Tab. \ref{tab:pm_abs} and Fig. \ref{fig:pm_abs}), while
the uncertainties properly take into account the substantial overlap in Gaia stars between
the two filters' samples.
However, it must be noted that the {\it Gaia} EDR3 proper-motion system itself may have an
unknown systematic error, i.e., offset from the inertial frame, 
at the location of Leo I.
One way to get a handle on the expected size of such an error is 
from a fit to the covariances within EDR3 as a function of angular separation, 
as for instance given by Eq.~1 of \citet{Li2021}.
This yields an estimate of 0.022 mas~yr$^{-1}$ for our field size.
We do not explicitly include this in our uncertainty budget and simply adopt
the {\it Gaia} EDR3 system as inertial. The reason for this is that we are not certain
of the behaviour of Eq.~1 of \citet{Li2021} over a field of view as small as ours and
because we will include in our final result information from ten galaxies.

\begin{figure}
\includegraphics[scale=0.45,angle=-90]{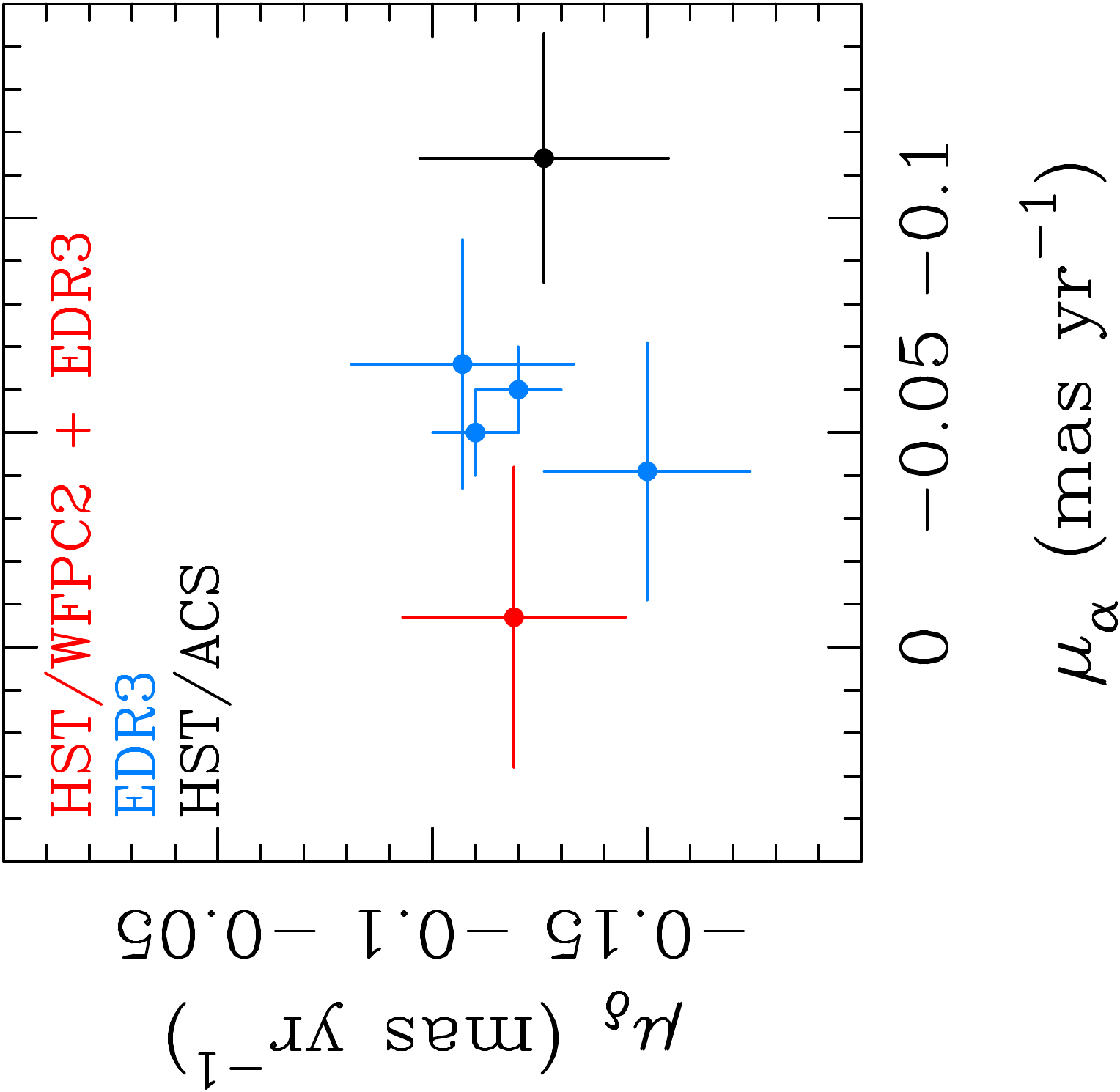}
\caption{Absolute proper-motion measurements of Leo~I: the galaxy-based HST/ACS 2013 one 
(black),
  the {\it Gaia} EDR3-only, large-area measurements (blue), and our
  current result (red). See also Tab. \ref{tab:leo1studies}. Note the scale of the plot which 
is
a zoomed-in version of Fig. \ref{fig:pm_status}.}
\label{fig:all-measurements}
\end{figure}

Finally, we combine our {\it Gaia} EDR3-based motion with that based on galaxies from Sec. 
\ref{subsec:galaxy-based},
doing once again a weighted average. Our final, combined result is:
$(\mu_{\alpha}, \mu_{\delta}) = (-0.007\pm0.035,-0.119\pm0.026)$ mas yr$^{-1}$. We display this 
final result in
Figure \ref{fig:all-measurements} together with the various large-area, {\it Gaia} EDR3-only 
measurements, and
the earlier HST/ACS galaxy-based result. We also list our final result
in Table \ref{tab:leo1studies}.

While all measurements agree in $\mu_{\delta}$ within errors, the recent {\it Gaia} EDR3-based 
results
--- including ours --- tend to have more positive values in $\mu_{\alpha}$ compared to the 
earlier HST/ACS result.

The question arises as to how best to combine the various measurements of Leo~I's absolute
proper motion, as given in Table \ref{tab:leo1studies} and shown in 
Figure \ref{fig:all-measurements}, if one desires the most useful value
for the purpose of an orbit analysis, for instance.
The {\it Gaia} DR2-based determinations are clearly superceded by equivalent 
{\it Gaia} EDR3-based ones.
Among the studies relying solely on {\it Gaia} EDR3 catalog proper motions of Leo~I members,
there is significant overlap and, hence, these are far from being independent measures.
These studies, which differ primarily in the methods used to determine membership in Leo~I
over the large area associated with the satellite galaxy,
include \citet{Battaglia2021,McConnachie2020} and \citet{Li2021}.
Among these three studies, only \citet{Li2021} takes into account systematic errors in
{\it Gaia} EDR3 using a covariance function built from values in \citet{Lindegren2021}.
This is an internally-based estimation of systematic errors, and the outcome is that
formal errors in the \citet{Li2021} study are $0.026 - 0.029$ mas yr$^{-1}$,
compared to those in the \citet{Battaglia2021} and \citet{McConnachie2020} studies
which are $0.01$ mas yr$^{-1}$ (see Table \ref{tab:leo1studies}). Incidentally,
the error estimates of \citet{Li2021} are in better agreement with
the error estimates of \citet{Martinez-Garcia2021} who include QSO's to perform
a local correction of the absolute proper motion system of {\it Gaia} EDR3.

Our recommendation is to choose one of these three \citep{Battaglia2021,McConnachie2020,Li2021}
measurements and combine it with the other, more 
independent measures, using a simple averaging and deriving the final uncertainty based on
the scatter of the individual measures.
This approach is justified by the near impossibility of disentangling any remaining
dependencies between the various studies or to adequately judge the level of unaccounted for
systematic effects that might be present.
We select \citet{Battaglia2021} as representative of the studies based on
EDR3 catalog measures of assumed Leo~I members, although either of the
other two would have been equally valid.
Next we include the determination by \citet{Martinez-Garcia2021}, 
in which a QSO-based correction to the local {\it Gaia} EDR3 zero point is made,
giving this measure a significant degree of independence, and a means to
keep systematic errors in check by incorporating external information
given by the QSOs.
Our own measure is included as it is focused on a tighter area near the center of Leo~I,
does not depend on {\it Gaia} EDR3 stars being members of the satellite, and is partially
based on external galaxies as well.

Finally, there is the ACS/WFC determination by \citet{sohn2013space} to consider.
As noted earlier in Sec. \ref{sec:litrev}, its value of $\mu_{\alpha}$ is rather discrepant
relative to subsequent {\it Gaia} EDR3-based determinations.
While it could be argued that it is only about $50\%$ more discrepant than our own measure,
we nonetheless choose to exclude it from the sample.
This is partially motivated by the exercise of the following section in which the
adopted absolute proper-motion is used to explore the pole of Leo~I's orbit.
A previous study of Leo~I's orbit by \citet{pawlowski2020milky} effectively adopted the
\citet{sohn2013space} proper-motion value by heavily weighting it.
We wish to contrast this by examining the effect of adopting a proper-motion value that
is predominantly {\it Gaia} EDR3-based and subsequently has a larger (positive)
value of $\mu_{\alpha}$ relative to the ACS/WFC determination.
The following Section will demonstrate the sensitivity of the orbit pole analysis to using
such an absolute proper-motion value.

The simple mean of the selected three measures of Leo~I's absolute proper motion, together
with the formal error from the standard deviation about the mean, is:
$(\mu_{\alpha}^{3meas}, \mu_{\delta}^{3meas}) = (-0.036\pm0.016,-0.130\pm0.010)$ mas yr$^{-1}$.
The estimated uncertainties correctly reflect the larger overall scatter of
the $\mu_{\alpha}$ component compared to that of $\mu_{\delta}$.

\section{Orbital Pole} \label{sec:pole}

A discussion of the orbit of Leo~I is beyond the scope of this paper. 
We will address, however, one aspect of the orbit, namely the location of the orbital pole 
which is of importance when analyzing the ensemble of orbital poles of Milky-Way satellites.
The clumpiness of the MW's most massive satellites' orbital poles is referred to as the 
Vast Polar Structure (VPOS), and indicates coherence
in the motion of these satellites.
\citet{pawlowski2020milky} determine the location of the
7 most-clumped satellite poles at Galactic coordinates
$(l,b)_{VPOS}^{PK20} = (179.5\arcdeg,-9.0\arcdeg)$.
This is for the ``combined''
sample, i.e., {\it Gaia} DR2 measurements and best available HST and other measurements (see 
their Table 3).

We calculate a new pole location for Leo~I 
based on the proper motion values obtained in Sec. \ref{sec:results}.
The input parameters for Leo~I are listed in Table \ref{tab:spacevelocity_input},
where the heliocentric distance
and line of sight velocity are from \citet{pawlowski2020milky} and references therein. 
Likewise,
in the calculation of the Galactic position $(X,Y,Z)$ and velocity
components $(V_x,V_y,V_z)$, we will
use the same solar constants as in \citet{pawlowski2020milky}, in order to
have a meaningful comparison.
These are: the local standard of rest (LSR) velocity $V_{LSR} = 239$ km s$^{-1}$, the solar
peculiar motion with respect to the LSR
$(V_{x,\odot},V_{y,\odot},V_{z,\odot}) = (11.10, 12.24, 7.25)$ km s$^{-1}$,
and the location of the Sun from the Galactic center $d_{\odot} = 8.3$ kpc.

\begin{deluxetable}{lrc}
    \caption{Orbit-Pole Input Parameters \label{tab:spacevelocity_input}}
    \tablehead{
        \colhead{Parameter} &
        \colhead{Value}
        }
    \startdata
        $l$ (degs) &
        225.986 \\
        $b$ (degs) &
        49.112 \\
        $D_{\sun}$(kpc)
        & $254 \pm 16$ 
        \\
        $V_{LOS} \,(km s^{-1})$
        & $282.5 \pm 0.1$ 
        \\
    \enddata
\end{deluxetable}

In Table \ref{tab:poles} we present the orbital pole calculations for three different values
of the absolute proper motion of Leo~I: the value determined in this study 
(Sec. \ref{sec:results}), the average of the selected {\it Gaia} EDR3 three most-independent measurements
(Sec. \ref{sec:results}), and the value used for Leo~I in  \citet{pawlowski2020milky}. The latter value
is dominated by the \citep{sohn2013space} ACS/HST measurement as error-based weights were
employed by \citet{pawlowski2020milky} in averaging the measurement of Leo~I, and the ACS/HST
measurement had the smallest ones.

Tab. \ref{tab:poles} lists these proper-motion determinations and the Galactic rest frame velocity components
with corresponding uncertainties in the first six columns. The next columns present 
the $(l,b)$ of the pole, the 
separation angle ($\Delta$)
between this pole and the location of the average of the
seven most-clumped satellite poles $(l,b)_{VPOS}^{PK20}$ from \citet{pawlowski2020milky}. 
Finally, in the last column
we list the standard deviation of the separation, which is determined in a Monte Carlo 
fashion
by generating 1000 realizations drawn from a Gaussian distribution of the errors in proper 
motions,
distance, and line-of-sight velocity. The scatter in the separation is dominated by the
proper-motion errors. There is an additional uncertainty of $16\arcdeg$ in the adopted value of
$(l,b)_{VPOS}^{PK20}$ \citep{pawlowski2020milky} to consider.

Leo~I's orbital pole as determined by \citet{pawlowski2020milky}
from the combined {\it Gaia} DR2 \citep{gaia2018gaia} and HST/ACS \citep{sohn2013space} 
measurements
is outside the clumped region of the orbital poles of the majority of the bright MW satellites
\citep[see also Fig. 1 in][]{pawlowski2020milky}. However, the new proper-motion
determinations place Leo~I's orbital pole in much better
agreement with the alignment of the other MW satellites (Tab. \ref{tab:poles}),
further reinforcing the coherence of the VPOS structure.
With the improved formal proper-motion uncertainties, the
pole determination is also better constrained.

\begin{deluxetable*}{rrr|rrr|rrrr}
  \tablecaption{Velocities, Orbital Pole Location and Separation from $(l,b)_{VPOS}^{PK20}$
    \label{tab:poles}}
\tablewidth{0pt}
\tablehead{
    {Solution} &
    \multicolumn{1}{c}{$\mu_{\alpha}$} &
    \multicolumn{1}{c}{$\mu_{\delta}$} &
    \multicolumn{1}{c}{$V_x$} &
    \multicolumn{1}{c}{$V_y$} &
    \multicolumn{1}{c}{$V_z$} &
    \multicolumn{1}{c}{$l_{p}$} &
    \multicolumn{1}{c}{$b_{p}$} &
    \multicolumn{1}{c}{$\Delta$} &
    \multicolumn{1}{c}{$\sigma_{\Delta}$}  \\
                 &
    (mas yr$^{-1}$) &
    (mas yr$^{-1}$) &
    (km s$^{-1}$) &
    (km s$^{-1}$) &
    (km s$^{-1}$) &
    ($\arcdeg$) &
    ($\arcdeg$) &
    ($\arcdeg$) &
    ($\arcdeg$) \\
    }
\startdata
this study & $-0.007 \pm 0.035$ &  $-0.119 \pm 0.026$  & $-67\pm32$ & $-7\pm32$ & $173\pm26$ & 156 &  -19 & 24 & 13  \\
ave. 3 meas. & $-0.036 \pm 0.016$ &  $-0.130 \pm 0.010$ & $-90\pm14$ & $-25\pm14$ & $148\pm13$ & 175 & -31 & 22 & 9   \\
PK20       & $-0.110 \pm 0.026$ &  $-0.116 \pm 0.025$  & $-168\pm27$ & $-25\pm27$ & $102\pm22$ & 251 & -39 & 70 & 14
\enddata
\end{deluxetable*}

\section{Summary} \label{sec:summary}

We determine the absolute proper motion of Leo~I using HST/WFPC2 exposures in F555W and F814W 
spanning
up to 10 years. The astrometry benefits from a new calibration of the WFPC2 camera 
\citep{casettiwfpc2}.
The absolute proper-motion zero point is based on 174 {\it Gaia}
EDR3 stars and 10 galaxies. We also include between $\sim 2000$ and $2300$ Leo~I stars in this 
determination.
Some of the advantages of our determination are: \\
$\bullet$ measuring a large number of Leo~I stars with precise relative
proper motions that allow us to cleanly separate Leo~I members and field stars; \\
$\bullet$ focusing on a small area near the center of light of the galaxy, 
thus avoiding potential problems with proper-motion gradients and the need to correct
for the center-of-mass motion; \\
$\bullet$ using an absolute proper-motion correction based on {\it Gaia} EDR3 stars,
but one that does {\it not} require them to be Leo~I members; \\
$\bullet$ providing also, as a check, a measurement based on ten background galaxies. \\

Our final result for the absolute proper-motion of Leo~I is:
$(\mu_{\alpha}, \mu_{\delta}) = (-0.007\pm0.035,-0.119\pm0.026)$ mas yr$^{-1}$.
We combine this result with that of two other recent measurements 
from the literature to obtain a multi-study estimate of:
$(\mu_{\alpha}^{3meas}, \mu_{\delta}^{3meas}) = (-0.036\pm0.016,-0.130\pm0.010)$ mas yr$^{-1}$.
An orbital-pole analysis shows that either of these two proper-motion values indicates
that Leo~I exhibits motion coherent with the VPOS, thus reinforcing the significance of this 
structure.

\acknowledgments

Support for program HST-AR-15632 was provided by NASA through a grant from
the Space Telescope Science Institute, which is operated by the Association of
Universities for Research in Astronomy, Inc.

This study has made use of data from the European Space Agency (ESA) mission {\it Gaia} 
(\url{https://www.cosmos.esa.int/gaia}),
processed by the Gaia Data Processing and Analysis Consortium (DPAC, 
\url{https://www.cosmos.esa.int/web/gaia/dpac/consortium}).
Funding for the DPAC has been provided by national institutions, in particular the institutions 
participating in
the {\it Gaia} Multilateral Agreement.

We are very grateful to Tony Sohn for making available the list of galaxies from their 2013 
study.

\vspace{5mm}
\facilities{{\it HST} (WFPC2), MAST, {\it Gaia}}

\newpage
\bibliography{ms}{}
\bibliographystyle{aasjournal}

\end{document}